Synaptic potentiation facilitates memory-like attractor dynamics in cultured *in vitro* hippocampal networks

Mark Niedringhaus[1*], Xin Chen[2*], Katherine Conant[1,3] and Rhonda Dzakpasu[1,2,4]

1. Interdisciplinary Program in Neuroscience, Georgetown University Medical Center
2. Department of Physics, Georgetown University
3. Department of Neuroscience, Georgetown University Medical Center
4. Department of Pharmacology and Physiology, Georgetown University Medical Center

*These authors contributed equally to this work.

**ABSTRACT**

Collective rhythmic dynamics from neurons is vital for cognitive functions such as memory formation but how neurons self-organize to produce such activity is not well understood. Attractor-based models have been successfully implemented as a theoretical framework for memory storage in networks of neurons. Activity-dependent modification of synaptic transmission is thought to be the physiological basis of learning and memory. The goal of this study is to demonstrate that using a pharmacological perturbation on *in vitro* networks of hippocampal neurons that has been shown to increase synaptic strength follows the dynamical postulates theorized by attractor models. We use a grid of extracellular electrodes to study changes in network activity after this perturbation and show that there is a persistent increase in overall spiking and bursting activity after treatment. This increase in activity appears to recruit more "errant" spikes into bursts. Lastly, phase plots indicate a conserved activity pattern suggesting that the network is operating in a stable dynamical state.

# I. INTRODUCTION

A major focus in dynamical neuroscience is identifying the neural patterns of activity that characterize human behavior as well as its surroundings. For example, it is thought that organized network activity in the form of synchronized depolarization is critical to cognitive processes such as attention and memory consolidation [1-6]. How neurons code the diversity of features in the environment and the assessment of the dynamic range of temporal responses when presented with external stimuli are some of the fundamental questions currently under investigation. However, an equally important question is how neurons self-organize into clusters or assemblies of coherent activity. These clusters of neural activity are thought to represent patterns that define different features within the external environment and the cluster constituents might change to reflect different environmental elements. For example, the same neurons may be involved in coding a particular color, e.g. yellow, but there is certainly a difference in response if a pattern of yellow in a green background characterizes a lion hiding among green leaves instead of dandelions. While it is thought that the timing between neurons or neural assemblies is involved, how neural signals cluster or self-organize within a circuit when presented with a given stimulus is largely unknown.

It is clear that at any given moment all aspects of the environment cannot be processed simultaneously. Additionally, features such as objects and events cannot be processed with equal weight. The brain must impose a selection protocol to evaluate information that is the most relevant to address the task at hand as well as to prepare for future events. This involves mechanisms of attention – selecting which neural correlates of the environment to manipulate or modulate – as well as memory, which involves registering and recording changes within these correlates for future recall. For example, the firing rates of visual cortical neurons of macaque monkeys are modulated when the animal attends to a particular stimulus [7-9]. *In vivo* recordings from these animals show that during attention, activity increases along with an increased coherence in firing rates. Therefore, it appears that attention is a modulator of neural dynamics as it prepares the network to enter into a particular state of elevated activity.

Once network activity has been modulated in response to a given stimulus, this activity may need to be preserved; a memory based upon this activity pattern may need to be created for future use and it is thought that a stimulus-dependent persistence in neural activity underlies active, i.e., working memory [10-12]. This learning rule, first postulated by Hebb stipulates that persistent activity is the result of strong recurrent excitatory connections between co-active neurons [10, 13]. The self-organization of activity displayed by these neurons accounts for the 'delay between stimulation and response, that seems so characteristic of thought' [10].

This neural correlate of memory has been incorporated into computational models of memory and it is believed that the dynamical correlate of working memory is the attractor state [11, 13-17]. Attractor models consist of highly interconnected networks of neurons that are capable of information storage. When there is an increase in connection strength within a population of model neurons, a persistent elevation in firing rates will result after a delay corresponding to the presentation of a cue. The increase in activity is based on Hebbian-like modifications in the strength of the connection, i.e., the synapse, and is compatible with spike-time dependent plasticity, STDP [18]. Experimental studies support the presence of attractors *in vivo* during hippocampal-dependent memory tasks [19, 20] and this led us to ask whether similar patterns of activity might be retained in networks of hippocampal neurons in the absence of an intact anatomical architecture. Our experiments assess the impact on network dynamics after applying a pharmacological treatment that modulates the strength of network connections or synapses.

Synapses are the chemical junctions between two neurons and a majority of excitatory synapses can be substantially influenced by structural changes in post-synaptic processes known as dendritic spines [21,22]. Neural activity can influence both spine size as well as the abundance of excitatory 2-amino-3-(5-methyl-3-oxo-1, 2-oxazol-4-yl) propanoic acid (AMPA) glutamate receptor subunits [22,29]. Changes in synaptic excitability can facilitate an increase in the concentration of post-synaptic neurotransmitter receptors. This will in turn influence action potential probability and the resulting firing rate within a network of neurons. These types of synaptic modulations have been observed in association with learning and memory and are thought to underlie the neural substrate of memory known as long-term potentiation, LTP [23-28].

LTP results from the increase in synaptic efficacy between neurons. It can be induced via high frequency electrical stimulation between pairs of neurons, or chemical stimulation and has been shown to last from several hours to many days [30, 31]. Mechanisms that have been shown to underlie LTP involve increases in dendritic spine size and its associated increase in the number of 2-amino-3-(5-methyl-3-oxo-1,2-oxazol-4-yl) propanoic acid (AMPA) receptors [29, 32, 33]. This increase in spine size causes an increase in connection strength, inducing a Hebbian-like synaptic modification leading to an increase in spiking activity. If a population of neurons is subjected to this modification, they can self-organize and cluster into active assemblies of elevated activity. If this activity persists, these assemblies might exhibit attractor dynamics.

LTP has been well studied between pairs of neurons within the hippocampus, specifically on synapses between the Schaffer collateral axons and apical dendrites of the CA1 pyramidal neurons [30, 34, 35]. This is a common neural region of LTP investigation since LTP is most reliably evoked in brain areas known to play a role in memory and learning [36]. However the impact on network dynamics due to the synaptic modifications modulated by LTP protocols has not been widely studied in experimental systems. Computational models have successfully incorporated the attractor paradigm as a mechanism through which information storage can be reliable invoked. Therefore, the goal of these experiments is to assess whether a synaptic perturbation that is thought to underlie the physiological basis of memory is characterized on the network level by the theoretical postulates of memory.

Consequently, this paper reports on the temporal network activity that arises when a pharmacological paradigm of LTP - chemical LTP – is introduced in cultured hippocampal neurons. By the use of the chemicals forskolin and rolipram, a large fraction of synapses in the network [41] can be potentiated via a detailed biochemical pathway that is believed to increase the AMPA receptor density [40-42]. This provides an advantage to potentiation via high frequency electrical stimulation between two neurons and is therefore a useful technique to facilitate synaptic potentiation resulting in an increase in the probability of neuronal spike generation in large populations such as cultured neural networks.

We use an array of extracellular electrodes, a multi-electrode array (MEA), to record spontaneous electrical activity when networks of hippocampal neurons have been pharmacologically perturbed. MEAs have been widely used to characterize dynamical activity from *in vitro* networks of neurons [43-47]. In addition, MEA studies that implement electrical stimulation protocols on *in vitro* networks of either hippocampal or cortical neurons have been established demonstrating precedence for an *in vitro* learning paradigm [48-52]. Lastly, an important temporal pattern found within developing *in vivo* circuits is the widespread prevalence of bursting activity [53-55]. Bursts are important during development as they facilitate normal functioning in developing neurons that in turn helps to create viable connections. We use young networks of cultured hippocampal neurons to study how a chemical LTP paradigm modulates network activity. We study network interactions at a time when the dynamics display a rich mix of vigorous bursting and spiking activity suggesting that these early periods are when the competition between spikes and bursts is at maximal levels. In our experiments, we show that network-wide firing rates increase but the variability in inter-spike intervals decrease. In addition, we show that the bursting frequency dramatically increases after chemical LTP evoking an elevation of network activity reminiscent of attractor dynamics. We suggest that the competition between synaptic inputs into the neurons, stimulated by the increased potentiation, results in the restructuring of the bursts as they form tightly compacted epochs of persistent activity, which may be indicative of an attractor basin of memory formation within the neural circuit.

## II. MATERIALS AND METHODS

### A. Cell Cultures

All experimental procedures were carried out in accordance with the Georgetown University Animal Care and Use Committee (GUACUC). Hippocampal tissue was extracted from embryonic day 18 Sprague-Dawley rats using a protocol modified from [56]. Briefly, the neural tissue was finely chopped and digested with 0.1% trypsin followed by mechanical trituration. Upon reaching a single cell suspension, 200,000 cells were added to multi-electrode arrays (MEA, Multi Channel Systems MCS GmbH, Reutlingen, Germany) that were previously treated with poly-d-lysine and laminin (Sigma, St. Louis, MO) resulting in an approximate density of 600 cells/mm$^2$. Cultures were maintained in Neuralbasal A medium with B27 (Invitrogen, Carlsbad, CA) with bi-weekly changes and kept in a humidified 5% $CO_2$ and 95% $O_2$ incubator at 37$^o$C.

### B. Electrophysiological recordings

We recorded all spontaneous electrical activity using a multi-electrode array. This MEA is composed of 59 titanium nitride electrodes, one reference electrode and four auxiliary analog channels each of which is 30 $\mu$m in diameter, arranged on an 8x8 square array. The inter-electrode spacing is 200 $\mu$m. Upon plating, the cells in suspension adhere to the silicon nitride substrate of the MEA and after seven days spontaneous electrical activity is detectable. We use the MEA1060 preamplifier and sample electrical activity at a 10kHz acquisition rate in order to allow the detection of multi-unit spikes. The data was digitized and stored on a Dell personal computer (Round Rock, TX). Possible exposure to contaminants and fluctuations in osmolality and pH were significantly reduced during the data acquisition period by the use of an MEA cover made of a hydrophobic membrane [57]. This membrane provides a tight seal, is semi-permeable to $CO_2$ and $O_2$ and is largely impermeable to water vapor. Experiments from at least three MEAs for each condition, including controls, were performed on a heated stage at 37$^o$C for at least 45 minutes at 14 days *in vitro* (14DIV), a time point during development in which the network displayed vigorous spontaneous electrical activity and for which network connectivity is well-established [58]. To ensure reproducibility of results across animals, all reported experimental groups were comprised of multiple cultures derived from multiple experimental preparations. Results obtained from cultures within and across different preparations were not significantly different.

### C. Pharmacological Induction of LTP

We used the pharmacological agents forskolin (50µM) and rolipram (100nM) to induce chemical LTP. Forskolin was dissolved in cellular media to a stock concentration of 50mM. Rolipram was dissolved in dimethyl sulfoxide (DMSO) to a stock concentration of 100µM. Both chemicals and DMSO were acquired from Sigma-Aldrich (St. Louis, MO).

We applied this chemical LTP treatment to the cultured hippocampal neurons on 14DIV. Initially, baseline electrical activity was recorded for 20 minutes on a heated stage at 37°C. To induce chemical LTP, 100µL of conditioned media was first removed from the MEA. Into this conditioned media, 1µL of each stock solution of forskolin and rolipram was diluted. The treated media was then slowly added back into the MEA. MEAs were returned to the stage and recordings resumed immediately lasting for at least 30 minutes. Results are presented for the period 20 minutes after recording.

To control for possible solvent effects as well as mechanical artifacts arising from the exchange of solutions, a series of MEA recordings were performed on cultures in which 1µL of DMSO was diluted into the conditioned media of another set of cultures prior to returning it to the MEA. Neither forskolin nor rolipram were added to these MEAs (vehicle experiments).

### D. Data Analysis

We removed low frequency components by high-pass filtering all traces at 200 Hz. Extracellularly recorded spikes, i.e., downward voltage deflections from baseline, were detected using a threshold algorithm from Offline Sorter (Plexon Inc., Dallas TX), which was calculated as a multiple of the standard deviation (-5σ) of the biological noise. We made no attempt to discriminate and sort spikes by electrode since the shape of a spike changes significantly during a burst due to changes in membrane excitability.

In addition, for this study we concentrate on network activity and the signal from each electrode suitably reflects these dynamics.

We used custom software written in MATLAB (The MathWorks, Natick, MA) to analyze dynamical activity in the cultured hippocampal networks. To investigate changes in overall network activity, we calculated the average firing rate, FR, over a binned (10-second binsize), five-minute window for each electrode within an MEA. Values are reported as averages ±SEM. We then calculated the ratio of firing rates after treatment with respect to baseline for both the chemical LTP experiments and the vehicle. Next, to obtain a measure of spiking regularity, we calculated the coefficient of variation, CV, defined as the following:

$$CV = \frac{\sigma \langle ISI \rangle}{\langle ISI \rangle}$$

where $\sigma$ is the standard deviation of the inter-spike interval (ISI) distribution.

Next, we investigated changes in a common temporal feature found in cultured networks, the burst, as it represents a collective network response. In our experiments, we analyzed bursts from each individual electrode. After the spike detection process described above each electrode has a resulting spike train, $\tau_{st}(t)$, expressed as:

$$\tau_{st}(t) = \sum_{n=1}^{N} \delta(t - t_n)$$

where N is defined to be the total number of spikes, $t_n$ is the time of the $n$th spike and $\delta(t)$ is a delta function that indicates a spike taking place at time $t = t_n$. The inter-spike interval between spike $n$ and spike $n$-1 (n >1) is:

$$\tau^{ISI}_n = t_n - t_{n-1}$$

For both the control and chemical LTP experiments, we define a burst from each electrode to consist of no less than four spikes with a maximum inter-spike interval (ISI) of 100 ms. Log histograms of the ISIs indicated that this corresponded to the cutoff of the first peak (fig. 2) in both conditions. Lastly, the burst durations, $\Delta_i$, are defined to be:

$$\Delta_i = t_{spike_{final}} - t_{spike_{initial}}$$

The final result of the burst identification process resulted in an M x N matrix where M corresponds to the electrode number and N's are the time stamps of the spikes within the bursts.

Lastly, we generated return maps of voltage activity to investigate the presence of nonlinear dynamical structures in the envelope of each bursting eposide before and after chemical LTP treatment. We low-pass filtered (10 Hz) each electrode and plotted $V_{i,t}$ vs. $V_{i,t+1}$ where $V_i$ is the voltage corresponding to electrode $i$ at time, $t$. A regularly repeating motif suggests the presence of a conserved activity pattern.

## III. RESULTS

Fig. 1 presents raster plots of spiking activity over a 20-second time window from the control hippocampal networks (fig. 1A) and the hippocampal networks 20 minutes after the application of chemical LTP (fig. 1B). One row in each panel corresponds to one electrode and in each row each small vertical tick mark is a detected spike. Below each raster plot is an expanded view of activity that shows a mix of bursts and single spikes. The raw voltage trace from a selected electrode is presented at the bottom. The control network exhibits bursts of a long duration. After chemical LTP, the bursts appear to cluster into tightly organized episodes of shortened duration and higher frequency.

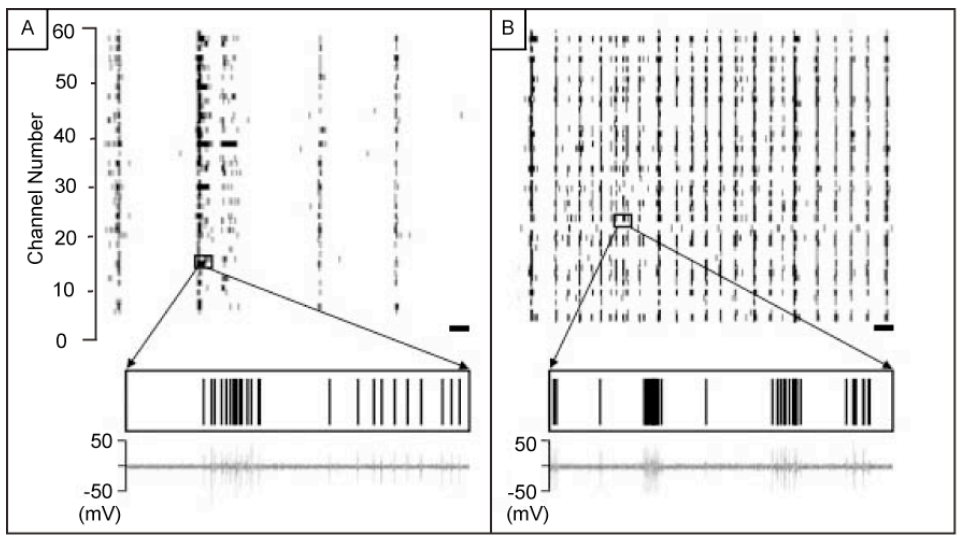

FIG. 1 Network spiking activity is increased after chemical LTP treatment. A) Raster plots of 20 seconds of spontaneous activity at 14 days *in vitro* from untreated cultured hippocampal networks. There is a large degree of activity with each electrode displaying bursting and spiking dynamics. B) Raster plots 20 minutes after application of chemical LTP. The expanded views show that the bursts increase in frequency and appear to shorten in duration. Scale bar=500 ms

We began our analysis by investigating changes in overall network activity. Fig. 2 is a log histogram of the inter-spike intervals from the chemical LTP and vehicle experiments showing that there is considerably more activity after chemical LTP. In addition to the large increase in activity, there is a leftward shift in the distribution. Within the short interval regime, usually corresponding to the spike intervals within bursts, is a well-defined peak around 5 ms embedded within a log normal-like distribution. In the longer interval regime there is a singular, pronounced peak near 10 seconds, an interval associated with being between bursts. The average ratio of firing rates (firing rate ratio after treatment relative to baseline) across the vehicle MEAs was 1.96 ±0.73 whereas the average ratio for the chemical LTP MEAs was 6.19 ±2.25. Fig. 3 highlights these differences in a spike count histogram using representative electrodes from the vehicle and chemical LTP treatments. There is an increase in spiking activity in the chemical LTP electrode while the activity in the electrode from the vehicle culture remains largely unchanged.

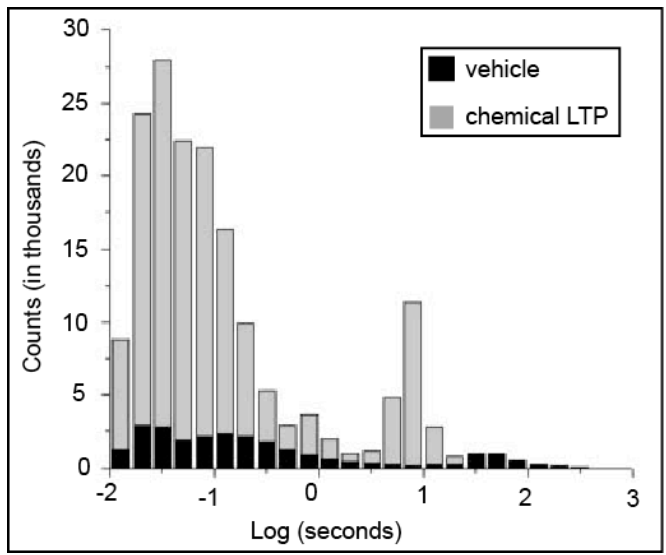

FIG. 2 There is a bimodal distribution of inter-spike intervals (ISI) after chemical LTP (grey bars). The first peak is clustered around short ISIs – this defines the intervals within the bursts whereas the second peak is near 10 second and corresponds to the interval between bursts.

Next we looked at the relationship between the aggregate number of spikes within a five-minute window before and 20 minutes after chemical LTP or vehicle treatment. Electrodes from all MEAs within each treatment were pooled and their spike counts are displayed on a log scale (fig. 4). The diagonal line represents y=x and therefore points falling on this line have no change in activity. Nearly all of the electrodes from the chemical LTP MEAs are above this line indicating an increase in activity, with a majority showing an increase of more than two orders of magnitude (fig.

4A). MEA vehicle experiments showed negligible change in the number of spikes (fig. 4B).

The profile of the time evolution of spiking activity in fig. 3 suggests that there is a change in the variability of inter-spike intervals (ISI) after chemical LTP. To address this, we calculated the coefficient of variation, CV, for all MEAs (fig. 5). There is a uniform decrease in the CV across all electrodes that experienced the chemical LTP treatment indicating that the variability in network activity was reduced. The change in the CV for the vehicle MEAs was negligible.

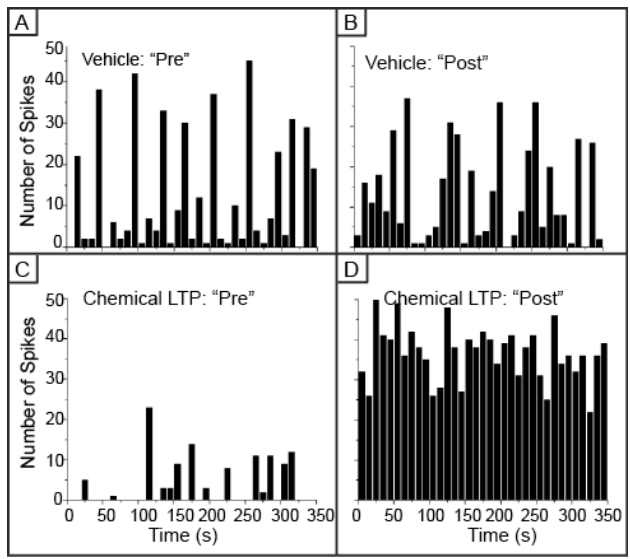

FIG. 3 Variability in spiking activity is reduced after chemical LTP. A,B) Spike count histograms from a representative electrode in the vehicle networks. There is robust but highly variable spiking activity. C) Spike count histogram in an electrode before chemical LTP. D) While the initial baseline is low in this example (C), the spike rate increases dramatically after chemical LTP treatment and the variability is low.

The increase in the firing rate and the decrease in variability of inter-spike intervals led us to ask how the chemical LTP treatment affected bursts, a subset of network activity. The burst, which is a tight barrage of spikes, is a dominant temporal motif in cultured networks, it is present in developing *in vivo* systems, and is believed to represent coordinated activity from neural assemblies [53-55]. It has been suggested that a burst may be more efficient to modulate information leaving a diminished role in information transmission for individual spikes [59-61]. If the bursts were positively impacted by the chemical LTP treatment, this would contribute to the increase in network regularity as seen in the reduction of the CV.

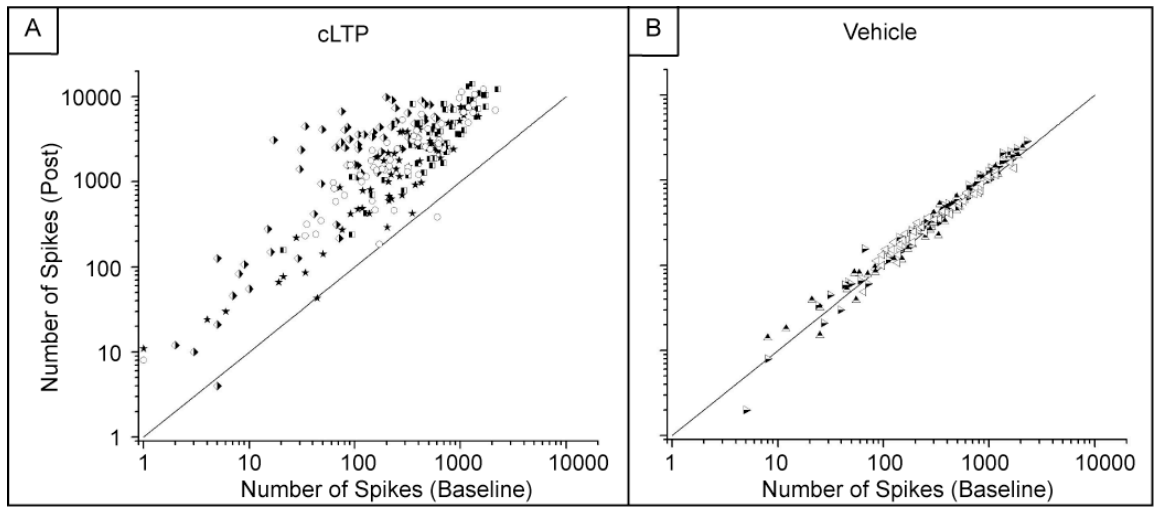

FIG. 4 There is a persistent increase in spiking activity after chemical LTP. A) Spike counts from all electrodes before and chemical LTP. Most electrodes have an increase in activity with a large cluster displaying an increase of at least two orders of magnitude. (one-way ANOVA, $p<10^{-9}$) B) Spike counts from the DMSO-treated MEAs show no increase in activity. (one-way ANOVA, $p<10^{-7}$). Each symbol corresponds to a different MEA. The diagonal line denotes y=x.

Fig. 6 presents the number of bursts and burst durations from the chemical LTP and vehicle MEAs. There is a significant increase in the number of bursts after chemical LTP and this increase clearly contributes to the increase in the overall firing rate within the network as seen in the raster plots of fig. 1. In the vehicle and pre-chemical LTP networks, the average number of bursts was approximately 2058±148 and 1564±429, respectively. However, the post-vehicle treatment increased the average number of bursts to approximately 2438±208 whereas 20 minutes after chemical LTP the average number of bursts increased to 10,300±2363. In addition, the burst durations decreased considerably after chemical LTP (fig. 6B). The average burst duration for the pre-chemical LTP MEAs was 140±18 ms and after treatment, 81±12 ms whereas the vehicle treatment the average was 130±3 before and 133±5 after treatment. This decrease in event duration suggests that the collective network activity contracted and experienced a re-organization into short episodes.

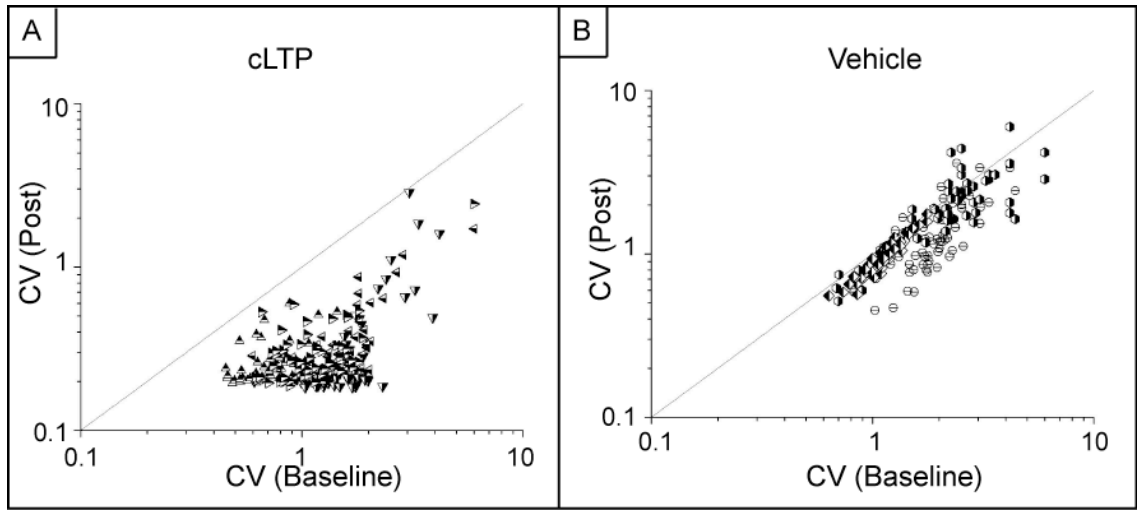

FIG. 5 The coefficient of variation (CV) of inter-spike intervals is reduced after chemical LTP. A) CV from the chemical LTP MEAs. There is an overall reduction in the CV indicating that variability in activity has been reduced. (one-way ANOVA, $p<10^{-5}$) B) MEAs treated with only DMSO show no change in the CV. (one-way ANOVA, $p<10^{-8}$) Each symbol corresponds to a different MEA. The diagonal line denotes y=x.

Bursts represent the collective network response to our pharmacological perturbation and only those spikes that participate within a burst are considered in the burst analyses. The raster plots of fig. 1 suggest that there may be a reduced number of spikes in between the bursts and therefore, we calculated the fraction of spikes not in bursts as a percent change from baseline. In the baselines of both the vehicle and chemical LTP experiment, approximately 20% of the spikes were not in bursts. However, there was a marked change after chemical LTP; this fraction decreased nearly 50% while the fraction in the vehicle fluctuated minimally. Chemical LTP appears to incorporate more of the "errant" spikes into bursts, leaving the inter-burst regions quiescent.

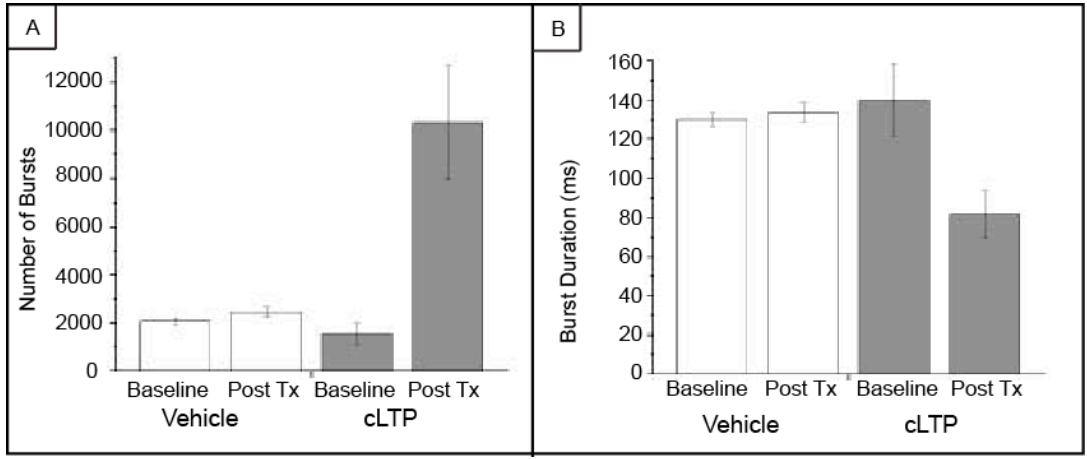

FIG. 6 Number of bursts and burst durations of spontaneous and evoked activity. A) The bursting activity significantly increases (one-way ANOVA, p=0.0003) after the application of chemical LTP, contributing to the overall increase in network firing rates as seen in fig. 2. B) The durations of the bursts decreases after chemical LTP (one-way ANOVA, p=0.0004).

Lastly, fig. 7 presents a representative return map of the low-pass filtered voltage for 10 seconds of activity from an electrode before (fig 7A) and after (fig 7B) chemical LTP. The baseline bursting activity pattern appears to be stable and the structure of each burst after chemical LTP is highly similar with a highly stereotyped spatiotemporal pattern.

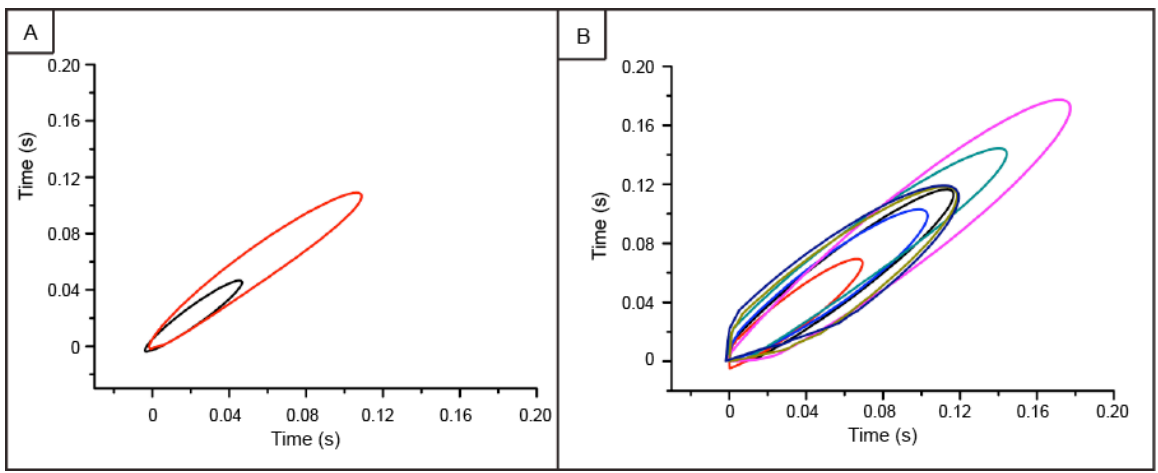

FIG. 7 (Color online) Conserved burst activity pattern is maintained after chemical LTP. A) Phase plot of bursts during 10 seconds of baseline activity. B) Phase plots of bursts during 10 second of activity after chemical LTP. Each motif repeats suggesting a preservation of an attractor state.

## IV. DISCUSSION

In these studies, we perform a global, biological manipulation that is believed to preferentially target a subset of structures residing on a small spatial scale – excitatory synapses ending on spines. We investigate the resulting dynamical effects on a large spatial scale - the network of cultured hippocampal neurons. The synapse was treated with a pharmacological paradigm that is known to increase the probability of action potential firing and quantified changes in spiking activity reflect the response from the network. This increase likelihood of firing is due to the Hebbian-like strengthening of synapses that might occur during the creation of a memory.

While the application of this drug cocktail may effect changes on the microscopic level other than synaptic modification, our results strongly indicate that synaptic perturbations can account for the observed modifications on the macroscopic level – overall network spiking activity. We suggest there may be two phenomena that could explain these changes. The chemical LTP treatment elevates network activity but the state remains stable. There is a major increase in overall network activity, as seen in the network firing rates. This is due to the increase in potentiation of a large fraction of synapses. This persistent activity due to an increase in connection strength has been theoretically described using attractor models.

In addition, there is a reduction in the coefficient of variation, CV, after chemical LTP. This reduction in the CV implies that the variability in the inter-spike intervals from the electrodes is reduced. The firing pattern becomes relatively constant with no large fluctuations of high activity. Regulation of neural activity must be preserved to prevent extremes in neural output – either hyperexcitability, which can lead to neurotoxic or neuropathological conditions, or insufficient excitation, which can cause the neuron to cease firing altogether. These regulatory mechanisms on the cellular level must also propagate to the network level in the form of circuit-stabilizing mechanisms and it has been suggested that appropriately modulated activity within a neural circuit could be maintained via the modulation of firing rates [62,63]. There may be a tuning range of firing rates over which the neural circuit operates most effectively. While it is too early after the treatment to assess long-term regulation of activity, our results suggest that the process of chemical LTP may facilitate the reduced variability of firing rates in the short term.

All of the firing rates from the electrodes increased dramatically after the chemical LTP treatment. However, the relative increase was not uniform across all electrodes and may be indicative of the different developmental stages of the neurons. These differences may also affect the ability of each neuron to respond to a synapse-strengthening perturbation. There is a small fraction of electrodes that displayed at least an eight-fold increase in multi-unit firing rate activity after treatment. This effect is further emphasized by the log scale presentation of spike counts produced by each electrode. As previously stated, we did not spike sort the data from these experiments. With our relatively low plating density, we rarely saw more than one unit per electrode (analysis not shown). We therefore introduce a possible scenario with the understanding that targeted biochemical assays are necessary to confirm our hypothesis. Chemical LTP modulates the neuron via several mechanisms and it will be the integrated effect that produces an increase in network-wide spiking activity. We focus, in this case, on one of these mechanisms and suggest that some of the neurons with this large firing rate increase are glutamatergic, i.e., excitatory, neurons with immature spines that responded with a vigorous spine expansion under chemical LTP induction. The spine expansion caused the firing rates of those cells to "catch up" to those of glutamatergic neurons with presumably more developed spines. This brought the previously immature cells within the range of the firing rates of the rest of the network. As a result, it appears from the dynamics within the network that all of the neurons, regardless of their initial developmental phase, had similar firing rates after treatment. Therefore, a striking network dynamical effect has materialized after chemical LTP in the reduced spiking variability. Chemical LTP has a differential effect on the increase in firing rates on clusters of neural assemblies, and these clusters may represent different information storing units.

Bursting activity in the network also displayed dramatic changes after synaptic potentiation. There is an increase in burst frequency, and the individual bursts are of a shorter duration. However, the decreased burst duration did not alter the shape of the attractor. The additional spikes generated by strengthening of the synapses need not have contributed to bursting activity and could have simply raised the background level of single spikes. Interestingly, not only did the burst frequency increase but there also was a large reduction in the fraction of spikes that are not participating in bursts accompanied by the preservation of the attractor profile. The large increase in the inter-spike interval histogram combined with the reduction of the number of spikes that do not participate in bursts suggest that the previously "errant" spikes were either recruited into existing bursts or, more likely, created new bursts with a shortened duration. It has been speculated that bursts may be more efficient at information processing within a neural circuit [59-61]. In these experiments, processing efficiency may represent information storage. The repeating spatiotemporal pattern observed in the return maps suggests that the system maintains a stable state of activity despite the persistent increase in activity. Lastly, the reduction in the coefficient of variation of inter-spike intervals suggests a more "regular" network temporal structure. These combined results demonstrate that synaptic potentiation evokes physiological events that restructure the burst profile. These restructured bursts represent the creation of a new functional entity that appears to facilitate information storage within the network.

## V. CONCLUSIONS

In conclusion, we show that a chemical paradigm that facilitates synaptic strengthening stimulates specific changes in network activity from cultured hippocampal neurons that are similar to results obtained from attractor-based computational models that describe memory storage. This demonstrates that cultured networks retain the essence of computational modeling as basic questions can be addressed on a reduced system. This system also preserves features of an *in vivo* model by using real neurons with their rich connectivity and complex patterns of activity. An applied stimulus to a neural system will influence its output, the spike. We asked the question, "Does a perturbation known to facilitate synaptic potentiation manifest as a dynamical correlate of memory on the global network scale?" While future studies are required for validation, the presence *in vitro* attractor dynamics in the form of persistent elevated activity suggests that fundamental principles of neural self-organization might be retained in the absence of anatomy.

Acknowledgements:

The authors are deeply grateful to Ernest Barreto, Joel Tabak and Maurizio Tomaiuolo for their valuable insights and suggestions.

M.N., X.C., and R.D. were supported by the Luce Foundation. K.C. was supported by a grant from the National Institutes of Health, National Institute on Aging (grant No. 029806).